\documentclass[11pt, oneside]{article}   	% use "amsart" instead of "article" for AMSLaTeX format
\usepackage{geometry}                		% See geometry.pdf to learn the layout options. There are lots.
\usepackage{graphicx}				% Use pdf, png, jpg, or eps§ with pdflatex; use eps in DVI mode
\usepackage{amssymb}
\usepackage{amsmath}
\usepackage[numbers]{natbib}
\usepackage{float}
\usepackage{xcolor}
\usepackage{color}
\usepackage{subcaption}

\providecommand{\keywords}[1]{\textbf{\textit{Keywords: \  }} #1}

\def\az{\mathcal{R}}

\def\palfRR{$f(\az^{\mu\nu}\az_{\mu\nu}) $ }
\def\cST{$c_{ST} $ }
\def\dL{the luminosity distance }

\date{}

\title{Observational effects of varying speed of light in quadratic gravity cosmological models}

\author{%
 \begin{tabular}{c}
  Azam Izadi \\  
  { Department of Physics, K. N. Toosi University of Technology, Tehran, Iran}\\
  aizadi@kntu.ac.ir\\  \tabularnewline
  Shadi Sajedi Shacker\\
  Department of Physics, University of Hamburg, Hamburger Sternwarte,\\
  Hamburg, Germany\\ 
  sshacker@hs.uni-hamburg.de\\ \tabularnewline
  Gonzalo J. Olmo\\
  Departamento de F\'{i}sica Te\'{o}rica and IFIC, Centro Mixto Universidad de Valencia\\ 
  - CSIC. Universidad de Valencia, Burjassot-46100, Valencia (Spain), \\
  and Departamento de F\'isica, Universidade Federal da Para\'\i ba,\\
  58051-900 Jo\~ao Pessoa, Para\'\i ba, Brazil\\
  gonzalo.olmo@uv.es\\ \tabularnewline
  Robi Banerjee\\
  Department of Physics, University of Hamburg, Hamburger Sternwarte,\\
  Hamburg, Germany\\ 
   rbanerjee@hs.uni-hamburg.de\\ \tabularnewline
\end{tabular}
}

\begin{document}
\maketitle
%\section{}
%\subsection{}

\begin{abstract}
We study different manifestations of the speed of light in theories of gravity where metric and connection are regarded as independent fields. We find that for a generic gravity theory in a frame with locally vanishing affine connection, the usual degeneracy between different manifestations of the speed of light is broken. In particular, the space--time causal structure constant ($c_{ST}$) may become variable in that local frame. For theories of the form $f(\az,\az^{\mu\nu} \az_{\mu\nu})$, this variation in $c_{ST}$ has an impact on the definition of the luminosity distance (and distance modulus), which can be used to confront the predictions of particular models against Supernovae type Ia (SN Ia) data. We carry out this test for a quadratic gravity model without cosmological constant assuming i) a constant speed of light and ii) a varying speed of light, and find that 
the latter scenario is favored by the data.\\
 \end{abstract}

\section{Introduction}

One of the major challenges faced by current cosmological models is the late-time accelerating expansion of the universe. 
According to the extensive observational evidence based on Type Ia supernovae (SN Ia) \cite{Riess, Perlmutter}, and standard rulers \cite{Shinji,palatini_gravity_theory}, our universe seems to be undergoing an accelerating expansion phase.
In order to explain this phenomenon, two general classes of models have been put forward. 
In the first class, the acceleration is attributed to new energy sources with repulsive gravitational properties, which is  dubbed dark energy. The second class of models seeks for self-accelerating solutions of the field equations by modifying the dynamics of general relativity (GR) on purely geometrical grounds or through the interpretation of cosmological observations from a different perspective \cite{uzan}. \\

The second approach offers a variety of alternatives to modify Einstein's theory of gravity. These fall into different categories such as scalar-tensor theories, tensor-vector-scalar theories, higher dimensional theories, theories with modified Lagrangians of the type $f(\az, \az^{\mu\nu}\az_{\mu\nu})$,  extensions that consider non-minimal matter-curvature couplings, and even more exotic alternatives. In addition, all these theories can be investigated in different variational scenarios, such as the (usual) metric approach, the Palatini/metric-affine formalism \cite{palatini_approach, fRgravity, Nojiri_unified, palatini_gravity_theory}, and also from a hybrid metric-Palatini perspective \cite{hybrid_metric_Palatini}.\\

When extensions of the standard model for gravity are considered, special attention should be paid to the origin and role of some fundamental constants. 
In particular, in metric-affine scenarios, where the metric and affine structures are a priori independent, it is necessary to clarify the different roles that the speed of light might play. 
This opens the possibility of distinguishing between several inequivalent manifestations which are degenerate in the framework of GR \cite{Ellis}. 
In this sense, one must distinguish between the following: 
\begin{itemize}
\item[$\bullet$] $c_{ST}$ : the space--time causal structure constant, which appears in the space-time line element and determines the local null cones.
\item[$\bullet$]  $c_{GW}$ : the velocity that appears in the equations that describe gravitational waves.
\item[$\bullet$]  $c_{EM}$ : the velocity that appears in electromagnetic waves equation.
\item[$\bullet$] $c_{E}$ : the gravity-matter coupling constant appearing on the right-hand side of Einstein's equations.
\end{itemize}
To the above, one could also add the {\it clock synchronization speed} $(c_C)$, which is the speed of the signal used to synchronize faraway clocks. 
Assuming Maxwell's electrodynamics and Einstein's gravity, the Newtonian limit requires $\frac{c_E}{c_0} = \frac{c_{GW}}{c_0} = \frac{c_{EM}}{c_0} = \frac{c_{ST}}{c_0} = 1$, 
where $c_0$ is a constant with dimensions, which we choose as $3 \times 10^8 m/s$ in MKS units.

It should be noted that beyond the Newtonian approximation there is no compelling reason to consider all the above manifestations of the speed of light to be the same in all theories and formalisms inasmuch as this quantity appears in many physical laws with different and {\it a priori} unrelated origins. \\

In the context of Palatini/metric-affine theories, the existence of a connection {\it a priori} independent of the metric puts forward that the $c_{ST}$ associated to the frames in which the metric is 
locally Minkowskian does not need to be the same as the quantity in a frame, in which the geodesics of the independent connection appear as straight lines. 
This is rather obvious for those cases in which there exists a nontrivial non-metricity tensor $Q_{\alpha\mu\nu}\equiv \nabla_\alpha^\Gamma g_{\mu\nu}$, where  $\nabla_\alpha^\Gamma$ 
denotes the covariant derivative of the independent connection. Given that $Q_{\alpha\mu\nu}$ is a tensor, the local coordinate transformation that brings the metric into Minkowskian form does not force 
the $Q_{\alpha\mu\nu}$ to vanish at the same time. As a consequence, though the Christoffel symbols ${\alpha\brace \mu\nu}$ of the metric vanish in those coordinates, the connection coefficients 
$\Gamma^\alpha_{\mu\nu}={\alpha\brace \mu\nu}-\frac{g^{\alpha\lambda}}{2}\left(Q_{\mu\lambda \nu}+Q_{\nu\lambda\mu}-Q_{\lambda \mu\nu}\right)$ do not, in general. 
Thus, locally $c_{ST}$ could be different from the propagation speed of light rays. 

In relation to this, the velocity of gravitational waves, $c_{GW}$ is another magnitude that deviates from the usual speed of light in theories beyond GR. In the case of Palatini theories whose Lagrangian is based on the metric and the Ricci tensor, it was recently shown that tensorial perturbations in cosmological scenarios satisfy an equation formally identical to that found in GR, in which the background variables represent the auxiliary metric associated to the independent connection \cite{Jimenez:2015caa}. This means that in the geometrical optics approximation, these gravitational waves follow the geodesics of $\Gamma^\alpha_{\mu\nu}$ rather than those of ${\alpha\brace \mu\nu}$. Though cosmological backgrounds of gravitational waves are not yet accessible to observation, the recent direct detection of aLIGO \cite{LIGO} opens a new window on the exploration of this new source of radiation and, in particular, for measuring and testing $\frac{c_{GW}}{c_{EM}}$  \cite{Collett,wave_fronts}.\\

The main purpose of this work, therefore, is to explore the impact of relaxing the universality of the different manifestations of the speed of light in the definition of 
 cosmological distances in the modified gravity scenario of Palatini extensions of GR. For concreteness, we will consider $f(\az)$ and $f(\az,\az_{(\mu\nu)}\az^{(\mu\nu)})$ theories with minimal matter couplings\footnote{It has been recently shown that these types of theories, which depend on the symmetrized Ricci tensor  $\az_{(\mu\nu)}$, have a projective invariance which makes the role of torsion trivial, being it possible to set the torsion to zero by a simple gauge choice \cite{Afonso}.}.     
We will comment on the different forms in which the speed of light enters in the equations that describe the background evolution of these theories and use observational data to study the effect of a varying speed of light on the supernovae luminosity distance. 
This analysis will allow us to determine if it is possible to find a modified gravity model with a varying speed of light able to fit the supernovae data and be in good agreement with the concordance model results without explicitly considering a cosmological constant or sources of dark energy. \\

\section{Speed of light in metric-affine theories}

In a very lucid work, Ellis and Uzan \cite{Ellis} presented different scenarios in which a quantity with dimensions of squared velocity appears and is typically interpreted as the same magnitude, namely, the speed of light. This assumption is nontrivial and comparable to the equivalence between inertial and gravitational mass. A deeper understanding of the latter led Einstein to formulate his now well accepted equivalence principle. A careful analysis of the different facets of the speed of light could lead also to a better understanding of the scenarios where it appears and their interrelations.\\

Though the relativistic formulation of physical laws was motivated by properties of the electromagnetic field, it is now well understood that the principle of relativity transcends Maxwell's theory and affects all the other known interactions. In this sense,  the limiting speed $c_{ST}$ that determines the causal structure of Minkowski space-time, $ds^2=c_{ST}^2dt^2-d\vec{x}^2$, needs not be the same as the speed $c_{EM}$ that appears in the equations for electromagnetic waves. The speed of light that appears on the right-hand side of Einsteins equations may also, in principle, be different from $c_{EM}$ and $c_{ST}$. These examples put forward that a careful analysis of the different facets typically attributed to the speed of light is necessary.   \\

The above discussion becomes particularly relevant in gravitational scenarios where the metric and affine structures of space-time are regarded as {a priori} independent. In such theories the local value of $c_{ST}$ may be subject to variation depending on how local observers are defined. This definition must be addressed carefully, as the dynamics of these theories may introduce operational subtleties not perceived in a purely kinematical analysis. To see this, let us consider first the kinematic description. In this approach, one may take the viewpoint that local measurements are performed by observers for which the {\it geodesics of the metric} $g_{\mu\nu}$ appear as straight lines around a given point. This simply requires performing a nonlinear coordinate transformation such that the original Christoffel symbols ${\alpha\brace \mu\nu}$ of the metric vanish. With an additional (linear) change of coordinates, $g_{\mu\nu}$ can be made Minkowskian at the chosen point, with the leading order 
corrections generated by the external matter fields being quadratic in the background curvature. If the total energy of the local gravitating system is small, then the Minkowskian approximation is valid locally in those coordinates. Alternatively, one may consider the family of local observers for which the {\it geodesics of the independent connection},  $\Gamma^\alpha_{\beta\gamma}$, are straight lines. In this case it is also possible to make the metric look locally Minkowskian by considering a local linear transformation that preserves the condition $\Gamma^\alpha_{\beta\gamma}=0$. Thus, it seems possible, in principle, to make the metric look Minkowskian at any desired point regardless of whether the connection is metric-compatible (Christoffel symbols of $g_{\mu\nu}$) or not. Obviously, the relation between the coordinates that make ${\alpha\brace \mu\nu}=0$ and those for which $\Gamma^\alpha_{\beta\gamma}=0$ is nonlinear, which implies that they have a certain relative acceleration (determined by the 
tensor $\Gamma^\alpha_{\beta\gamma}-{\alpha\brace \beta\gamma}$). In other words,  if one picks up a frame where only the Christoffel symbol of the metric is zero at some point, then the full Riemann tensor would not be zero at that point.\\

Now, the point is that once the dynamical relation between $g_{\mu\nu}$ and $\Gamma^\alpha_{\beta\gamma}$ is obtained, it is not immediate to guarantee that the Minkowskian condition can always be satisfied locally for $g_{\mu\nu}$. An explanation is in order. In these theories, one typically finds that $\Gamma^\alpha_{\beta\gamma}$ can be written as the Christoffel symbols of an auxiliary metric $h_{\mu\nu}$. This $h_{\mu\nu}$ is governed by a set of equations of the form ${\az_\mu}^\nu(h)={\tau_\mu}^\nu$, where ${\az_\mu}^\nu(h)=R_{\mu\alpha} h^{\alpha\nu}$, $\az_{\mu\alpha}$ is the Ricci tensor of $h_{\mu\nu}$, and ${\tau_\mu}^\nu$ represents an effective stress-energy tensor which is completely determined by $h_{\mu\nu}$ and the matter fields. This implies that locally (or in the absence of external gravitational fields) any departure of $h_{\mu\nu}$ from $\eta_{\mu\nu}$ is determined by an integration over the local matter sources, i.e., by the total energy of the local system (similarly as in GR). The 
point is that the relation between $h_{\mu\nu}$ and $g_{\mu\nu}$ depends on the local stress-energy {\it density} via a transformation of the form $g_{\mu\nu}={{\Omega^{-1}}_\mu}^\alpha h_{\alpha\nu}$, with ${{\Omega^{-1}}_\mu}^\alpha$ specified by the stress-energy tensor of the local sources. For $f(\az)$ theories, for instance,  ${{\Omega^{-1}}_\mu}^\alpha=\frac{1}{f_\az(T)}{\delta_\mu}^\alpha$, where $f_\az=df/d\az$ and the dependence on the trace $T$ of the stress-energy tensor is determined by the equation $\az f_\az - 2f(\az)=\kappa^2T$. Thus,  if the total energy-momentum of a local system is small but its stress-energy {\it density distribution} is not negligible, one could only justify that $h_{\mu\nu}\approx \eta_{\mu\nu}$ but $g_{\mu\nu}\approx \frac{1}{f_\az(T)} \eta_{\mu\nu}$. This property has been used to rule out models with inverse curvature terms due to nontrivial effects on atomic systems \cite{Olmo:2008ye,Olmo:2006zu,Flanagan:2003rb}. Obviously, at every single point of the local system 
(in space and in time) one could find a coordinate transformation that removes the density dependence from the metric at that instant and location. However, the notion of observer as a point-like, structureless entity is not valid in this context. For extended sources with fluctuating stress-energy tensor or for regions crossed by different radiation fields, such a local (point-like, structureless) and instantaneous choice of coordinates is of no use. One can thus, at most, screen the external effects of gravity by the choice of a frame in which $h_{\mu\nu}$ appears as Minkowskian. The effects of the local stress-energy density, however, cannot be eliminated in that way.  \\

Turning back to the definition of $c_{ST}$, one finds that in the case of $f(\az)$ theories, $c_{ST}$ is insensitive to the form of the function $f_\az(T)$ because the two metrics are conformally related. In other metric-affine theories, such as in the case of quadratic gravity,  $\az+\alpha \az^2+\beta \az_{(\mu\nu)} \az^{(\mu\nu)}$, or in Born-Infeld inspired gravity models, the deformation matrix ${{\Omega^{-1}}_\mu}^\alpha$ has two (or more) different eigenvalues, which induces nontrivial modifications in $c_{ST}$. As discussed in \cite{Izadi_1, Izadi_2} for FLRW cosmological models driven by perfect fluids in gravity theories of the form  $\az+f( \az^{\mu\nu} \az_{\mu\nu})$, one finds ${{\Omega^{-1}}_\mu}^\alpha=\text{diag}(1/(\lambda \omega),1/\lambda,1/\lambda,1/\lambda)$, where $\lambda$ and $\omega$ are functions of the fluid pressure and energy density, which are functions of cosmic time. In the local frame $e^\mu_a$ in which $h_{\mu\nu}e^\mu_a e^\nu_b=\eta_{ab}$, one has $g_{\mu\nu}e^\mu_a e^\nu_
b=\text{diag}(1/(\lambda \omega),-1/\lambda,-1/\lambda,-1/\lambda)$. As a result, the line element $ds^2=g_{\mu\nu}dx^\mu dx^\nu$ becomes $ds^2=c_0^2dt^2-d\vec{x}^2$ in its own local frame but turns into $ds^2=c_0^2/(\lambda \omega)dt^2-d\vec{x}^2/\lambda$ in the local frame of $h_{\mu\nu}$. Thus, in the $h$-frame, we have $c_{ST}^2=c_0^2/\omega$. This puts forward that $c_{ST}$ may be subject to a time variation induced by the evolution of the cosmic expansion and modulated by the specific form of the gravity Lagrangian.  

% An observer in a local inertial frame in curved space-time can compare tensors and vectors at neighboring events, just as he would in flat space-time. [MTW, page 208] 
% This statement associates local inertial frames with frames where the connection vanishes. 

\section{Dynamics of Palatini $f(\az)$ and $f(\az, \az^{\mu\nu} \az_{\mu\nu})$  theories}

For the sake of concreteness, in this section we study two well-known families of theories whose Lagrangians are arbitrary functions of the form $f(\az)$ and $f(\az, \az^{\mu\nu} \az_{\mu\nu})$ (from now on we assume symmetrization of the indices in the Ricci tensor). Their field equations will be derived assuming that metric and connection are independent. For simplicity in the discussion, we will consider $c_E=c_0$ so that $\kappa=\frac{8 \pi G}{c_0^4}$.\\

The Lagrangian density of the non-linear Ricci scalar gravity model is chosen to be an arbitrary function of the scalar curvature, 
\begin{equation}
\mathcal{L} = \frac{1}{2\kappa} f(\az[g,\Gamma]) + \mathcal{L}_m
\label{action}
\end{equation}
\noindent in which $\kappa=8\pi G/c_0^4$, and the matter Lagrangian density, $\mathcal{L}_m$, does not depend on  the connection for simplicity. Variation of the action leads to\footnote{For a precise derivation of these equations including torsion, see \cite{Afonso}}
\begin{equation}
f'(\az)\az_{\mu\nu}-\frac{1}{2}f(\az)g_{\mu\nu}=\kappa T_{\mu\nu} \ ,
\label{eq1}
\end{equation}
whose trace with respect to the metric yields the algebraic relation $\az f_\az-2f=\kappa T$. The connection equation can be simplified to 
\begin{equation}\label{eq01}
\nabla_\alpha^\Gamma \left (\sqrt{-g}f'g^{\mu\nu}\right )=0 \ ,
\end{equation}
where $\nabla_\mu^\Gamma X^{\nu}= \partial_{\mu}X^{\nu} + \Gamma^{\nu}_{\alpha\mu}X^{\alpha}$.
It is easy to verify that the connection that solves this last equation coincides with the Levi-Civita connection of a metric $h_{\mu\nu}=f'g_{\mu\nu}$, such that
\begin{equation}\label{ggama}
\Gamma^{\alpha}_{\mu\nu}= {\alpha\brace \mu\nu}+ \gamma^{\alpha}_{\mu\nu}={\alpha\brace \mu\nu}+\frac{1}{2f'}\left [
2\delta^{\alpha}_{(\mu}\partial_{\nu)}f'-g_{\mu\nu}g^{\alpha\beta}
\partial_{\beta} f'\right ] \ .
\end{equation}
\noindent For the case of $f (\az, \az_{\mu\nu}\az^{\mu\nu})=\az+f(\az^{\mu\nu}\az_{\mu\nu})$ theories, it can be shown that variation of the action with respect to the metric and the independent affine connection leads to \cite{Afonso}
\begin{equation}
\az_{\mu\nu} + 2F\az_{\mu}^{\alpha}\az_{\nu\alpha} - \frac{1}{2}g_{\mu\nu} [\az + f] = \kappa T_{\mu\nu}
\label{mee}
\end{equation}
\begin{equation}
\nabla_{\sigma}^\Gamma \left[\sqrt{-g} (g^{\mu\nu} + 2Fg^{\mu\alpha}\az_{\alpha\beta}g^{\nu\beta})\right] = 0 \ .
\label{ce}
\end{equation} 
After ADM decomposition, the energy--momentum tensor can be written as below  \cite{Izadi_1, Mota}:  
\begin{equation}\label{Tmunu}
T_{\mu\nu}=\rho u_\mu u_\nu + 2q _{(\mu} u_{\nu)} - p \tilde{g}_{\mu\nu} + \pi_{\mu\nu},
\end{equation}
\noindent in which $u^\mu = \frac{dx^\mu}{d\tau} $ is the 4-velocity normalized as $ u^\mu u_\mu=1$, $ \tilde{g}_{\mu\nu} = g_{\mu\nu} - u_\mu u_\nu$ which determines the orthogonal metric properties of observers moving with 4-velocity $ u^\mu$, 
$ \pi_{\mu\nu}= \tilde{g}^\alpha_\mu \tilde{g}^\beta_\nu T_{\alpha \beta}$ is the projected symmetric trace free anisotropic pressure, 
$ \rho = T_{\mu\nu} u^\mu u^\nu$ is the relativistic energy density relative to $ u^\mu$, 
$ q_\mu = \tilde{g}^\alpha_\mu u^\beta T_{\beta\alpha}$ is the relativistic momentum density, 
and $ p = \frac{-1}{3} \tilde{g}^{\mu\nu}T_{\mu\nu}$ is the isotropic pressure. \\
Since the modified Einstein equation (\ref{mee}) is an algebraic equation, to go further let us write the symmetric Ricci tensor $\az_{\mu\nu}$ in a general way as  
\begin{equation} \label{Rmunu}
 \az_{\mu\nu} = \Delta u_\mu u_\nu + \Xi \tilde{g}_{\mu\nu} + 2 u_{(\mu}\gamma_{\nu)} + \varSigma_{\mu\nu}.
\end{equation}
Substituting  \eqref{Tmunu} and \eqref{Rmunu} in the modified Einstein equation (\ref{mee}) leads to the four below equations:   
\begin{equation}
\Delta + 2F\Delta^{2} - \frac{1}{2}(\Delta + 3\Xi + f) = \kappa\rho c_E^2
\label{1}
\end{equation}
\begin{equation}
\Xi +2F\Xi^{2} - \frac{1}{2}(\Delta + 3\Xi + f) = -\kappa p
\label{2}
\end{equation}
\begin{equation}
[1 + 2F(\Delta + \Xi)] \Upsilon_\mu = \kappa q_\mu
\label{eq3}
\end{equation}
\begin{equation}
(1 + 4F\Xi) \Sigma_{\mu\nu} = \kappa\pi_{\mu\nu}
\label{eq4}
\end{equation}
where, recall, $f$ and $F$ are functions of $\az^{\mu\nu}\az_{\mu\nu} = \Delta^{2} + 3\Xi^{2}$. For FLRW cosmological background $q^\mu = \pi^{\mu\nu} = 0$ and accordingly we have $\Upsilon_\mu = \Sigma_{\mu\nu} = 0$. Therefore, given the specified form of $f(\az_{\mu\nu}\az^{\mu\nu})$ and the values of the density $\rho$ and the pressure $p$, all unknown coefficients in the above equations can be determined (at least numerically).\\

%The affine connection and the Christoffel symbol of the physical metric are related by the tensor $\gamma^{\alpha}_{\mu\nu}$, \cite{Izadi_1, Mota}, 
%\begin{equation}
%\Gamma^{\alpha}_{\mu\nu}={\alpha\brace \mu\nu}+ \gamma^{\alpha}_{\mu\nu}
%\label{18}
%\end{equation}
%\noindent The background evolution of the general Ricci squared gravity has been studied until here at the first perturbation level. As a special example, we will investigate a specific family of these theories with $f(\az^{\mu\nu} \az_{\mu\nu})=F \az^{\mu\nu} \az_{\mu\nu}$, and we will constrain the allowed parameter space by using the SN Ia luminosity distances.

\section{Speed of light in $f(\az, \az^{\mu\nu} \az_{\mu\nu})$ theories}
Let us now investigate the properties of the speed of light in Palatini theories of the form $\az + f(\az^{\mu\nu} \az_{\mu\nu})$.
%All aspects of the speed of light coincide with the $c_0$ in the standard gravity in the first local frame, in which the metric is locally Minkowskian. But this is not the case in the second local frame, in which the auxiliary metric is locally flat and the affine connection is locally zero.
As shown in \cite{Izadi_1} and discussed above, in the local frame in which the affine connection is locally vanishing, the degeneracy of different aspects of the speed of light is broken.
%we normally consider for the speed of light in the MKS units.  
In particular, for nonlinear Ricci squared cosmological models, the causal structure constant in the local frame where the independent connection vanishes takes the form
\begin{equation}\label{cST}
 c_{ST} = \frac{c_0}{\sqrt{\omega}}.
\end{equation}
In order to find \cST, one needs to know $\omega$ as a function of the density $\rho$:
\begin{equation} \omega=\frac{1+2F\Xi}{1+2F\Delta}. \end{equation}
Here $F$ is the derivative of $f(\az^{\mu\nu}\az_{\mu\nu})$ with respect to $\az^{\mu\nu}\az_{\mu\nu}$ and $\Xi$ and $\Delta$ can be obtained from the equations below:

\begin{equation}\label{delta}
\Delta+2F\Delta^2 -\frac{1}{2}(\Delta+3\Xi+f)=\kappa\rho c_E^2 
\end{equation} 
\begin{equation}\label{xi}
\Xi+2F\Xi^2 -\frac{1}{2}(\Delta+3\Xi+f)=-\kappa p.
\end{equation} 
Given the values of $\rho$ and $p$, the $\Delta$ and $\Xi$ coefficients and $\az^{\mu\nu}\az_{\mu\nu} = \Delta^{2} + 3\Xi^{2}$ can be obtained from the above equations. \\

For the present model, the Hubble parameter is obtained by solving the modified Einstein field equation (\ref{mee}) in FLRW background.
\begin{equation}
(H+ \frac{\dot{\lambda}}{2\lambda})^2 = \frac{c_0^2}{6}(\Delta - 3\omega \Xi).
\end{equation}
Since $\lambda$ is a function of $\rho$, consequently the above equation can be expressed as 
\begin{equation}\label{Hmod}
H^2 (1- s \frac{\lambda^\prime \rho}{2\lambda} )^2 = \frac{c_0^2}{6}(\Delta - 3\omega \Xi),
\end{equation}
in which $c_0=c_E$, $\lambda = \sqrt{(1+2F\Delta)(1+2F\Xi)}$, $\lambda^\prime =\frac{\partial \lambda}{\partial \rho}$,  and $s= 3$, $4$ for matter and radiation eras, respectively \cite{Mota}.  \\
For our specified model, the quantity $\omega$ can be obtained in the different cosmological eras. 
For the matter dominated era,
\begin{equation}
\label{omega}
 \frac{1}{\sqrt{\omega}} =
\sqrt{\frac{1+\kappa F \rho_m^0 c_0^2  (1+z)^3}{1-\kappa F \rho_m^0 c_0^2  (1+z)^3}} 
\end{equation}
And $\omega\rightarrow 1$ for the present/future de Sitter era. It is important to mention that \cST has been larger than its present value in the past \cite{Izadi_1}. \\

As mentioned in the introduction, from the analysis of the propagation of gravitational waves in metric-affine theories based on the Ricci tensor \cite{Jimenez:2015caa}, one finds that gravitational waves in the geometrical optics approximation, propagate along the lines associated to the independent connection $\Gamma^\alpha_{\beta\gamma}$. This is a convincing motivation for us to investigate how the equations describing cosmological observables would be modified considering that \cST has not always been $c_0$.\\
Given that in these theories the potential varying effects are due to the presence of local stress-energy densities,  in vacuum space-times such as in the Schwarzschild case, there is no potential conflict because both the independent affine connection and the Christoffel symbols of $g_{\mu\nu}$ coincide. Therefore, for local tests outside the matter distribution ($T_{\mu\nu}=0$) no variation can be detected. However, in regions where the matter density is not zero, such as in cosmological models, the choice of local frame becomes nontrivial in metric-affine theories.

\section{Luminosity distance, $c_{ST}$, and  Supernovae type Ia}
\label{sec:Friedmannology}

In standard cosmology, the luminosity distance, $d_L$, is defined in such a way as to preserve the Euclidean inverse-square law for the weakening of light with distance from a point source \cite{Shinji}:
\begin{equation}\label{eq:d_Lusual}
 d_L = f(\chi)\sqrt{\frac{L_s}{L_o}}=\frac{(1+z)c}{H_0\sqrt{\Omega^0_k}} \sinh \left(\frac{\sqrt{\Omega^0_k}}{c} \int \frac{c dz}{E(z)}\right) \ ,
\end{equation}
in which\footnote{Note that in terms of wavelength the observational redshift can be defined as
\begin{equation}\label{zprime}
1+z_{obs} =\frac{a_0}{a(t)}=\frac{c_0 dt_o}{c_{ST} dt} = \frac{\lambda_{obs}}{\lambda_{em}} \ ,
\end{equation}
where $\lambda_{obs}$ and $\lambda_{em}$ are the observed wavelength and the emitted wavelength, respectively.} $\sqrt{\frac{L_s}{L_o}} = (\frac{\Delta \nu_1}{\Delta \nu_0}) = (1+z)$,  
$f(\chi) = \frac{c}{H_0\sqrt{\Omega^0_k}} \sinh \Big(\frac{\sqrt{\Omega^0_k}}{c} \int \frac{c dz}{E(z)}\Big)$, $E(z)=H(z)/H_0$, and $H(z)$ is the Hubble function, with $H_0$ being its observed value today. Here the density parameter is defined as $\Omega_k^0=\frac{-kc^2}{(a_0H_0)^2}$
and $k$ denotes the curvature of the spatial sections: open when $k=-1$, flat for $k=0$, or closed if $k=+1$. 
From an observational point of view, \dL is typically written as \cite{Coles}
\begin{equation}\label{dL_observation}
 d_L = 10^{1+\mu/5} pc,
\end{equation}
in which $\mu= m-M $ is the distance modulus, $m$ is the apparent magnitude, and $M$ is the absolute magnitude.\\

When the degeneracy between the different manifestations of the speed of light is broken, the definition of the luminosity distance must be reconsidered. Due to the fact that the definition of distance and 
specifically the luminosity distance comes from the definition of the line element, therefore, we should replace $c$ in the equation (\ref{eq:d_Lusual}) by $c_{ST}$. 
Accordingly, Eq.(\ref{eq:d_Lusual}) must  be replaced by 
\begin{equation}\label{newdl} 
d_L=\frac{(1+z) c_{ST}}{H_0\sqrt{\Omega_k^0}}\sinh\left(\frac{\sqrt{\Omega_k^0}}{c_{ST}}\int_{0}^{z}{c_{ST}\frac{d {z}}{E({z})}}\right),
\end{equation}
which for a flat universe boils down to
\begin{equation}d_L=(1+z)\int_{0}^{z}{c_{ST}\frac{d {z}}{H}} \ ,
\label{dL_VSL}
\end{equation}
with $H$ given by equation \eqref{Hmod}. 

At this moment, back to the section $2$, choosing an appropriate local frame plays a key role. If we choose a frame, in which the Christoffel symbol of the metric is locally vanishing, there is no variation in the speed of light, which implies $c_{ST} = c_0$. Hence, the above equation for a flat universe takes the standard form
\begin{equation}d_L=(1+z)\int_{0}^{z}{c_{0}\frac{d {z}}{H}} \ ,
\label{dL_MOG}
\end{equation}
where $c_{ST}$ has just been replaced by $c_0$. \\

When the local frame is taken as that in which the independent affine connection is locally vanishing, the story is different because this is the frame in which the degeneracy of the different manifestations of the speed of light is automatically broken. Even so, for Palatini $f(\az)$ models we showed before that  in the local frame $c_{ST}$ coincides with $c_0$, implying no new observational effects as far as light propagation measurements are concerned. 
On the contrary, when \palfRR corrections are considered, the varying character of $c_{ST}$ leads to potentially observable effects. 
To explore them, we consider as an example a Ricci-squared correction to the GR Lagrangian of the form 
\begin{equation}\label{eq:model}
f(\az_{\mu\nu}\az^{\mu\nu})=F \az_{\mu\nu}\az^{\mu\nu} \ ,
\end{equation}
%with $F = l_P^2 \Gamma$, where $l_P$ is the Planck length and $\Gamma=10^{120}\gamma$.  Therefore, $ \gamma$ will be the free parameter of the model.  
with free parameter $F$, which has the  dimension of $L^{2}$. For numerical convenience,  we rewrite $F=\gamma \Gamma$, in which $\Gamma=2.743 \times 10^5\ Mpc^2$.
Our purpose now is to confront recent SN Ia data with the above model (\ref{eq:model}) to see how, from a statistical perspective, a varying speed of light cosmology performs as compared to the case of a strictly constant  $c_{ST}$. 
The fits will also be compared with the standard $\Lambda$CDM model, for which we take $\Omega^0_m= 0.328$ and also $H_0= 65.1\ km s^{-1} Mpc^{-1}$, where the zero indices 
denote the value in the present era \citep{Planck_VFC}. 
For this purpose, the $580$ measurements on SN Ia luminosity distance in the union 2.1 data set are used \citep{union}. 
%The results are shown in Figure \ref{fig_dL} for different $\gamma$ values. \\
%\\As a more general correction, the model $f(\az+\az_{\mu\nu}\az^{\mu\nu}) = aR^2 + F\az_{\mu\nu}\az^{\mu\nu}$ with a new free parameter $a$, will be studied in a future work.\\

\subsection{Confronting models with observations}\label{sec:confrontation}

For the confrontation of models with observations  we use the $\chi^2$ minimization test. 
The observed distance modulus $\mu_{obs}=5\ logd_L+25$ and its uncertainty, $\Delta \mu_{obs}$, for individual SN Ia is supplied by the measurements \citep{union}. Therefore, the standard $\chi^2$ minimization, which is defined by  
\begin{equation}
\chi^2= \sum_{i=1}^{580} \frac{(\mu^i_{obs}-\mu^i)^2}{(\Delta \mu^i_{obs} )^2}
\end{equation}
can be implemented, where $\mu$ is the theoretically predicted distance modulus for a model. \\

We study the following models: (i) The concordance model ($\Omega^0_m= 0.329$ and $\Omega^0_\Lambda=0.671$ \citep{Planck_VFC}) (ii) The Modified Einstein-de Sitter (MEdS) model ($\Omega^0_m= 1$ together with \eqref{Hmod} and \eqref{dL_MOG}) and the VSL model ($\Omega^0_m= 1$ together with \eqref{Hmod} and \eqref{dL_VSL}).
The last two models allow to compare the difference between simply having a modified Lagrangian and a modified Lagrangian with a broken degeneracy between \cST\ and $c_0$.

For the concordance model, $\chi^2$ minimization leads to $\chi^2_{CM} = 936.801$. 
For the MEdS model assuming a constant speed of light [see Eq.(\ref{dL_MOG})], we find the minimum $\chi^2$ at $\chi^2_{min,MEdS}=922.712$ for $\gamma=1.107$. 
Within the $68 \%$ confidence level, this leads to $62.8<H_0<63$, which is at odds with Planck observations  ($63.2<H_0<66.8$ at the $68 \%$ confidence level \cite{Planck_VFC}). 
So we conclude here that within the $68 \%$ confidence level, MEdS model is inconsistent with Planck data for $H_0$ value.
When the varying speed of light relation (\ref{dL_VSL}) is imposed on the model  (\ref{eq:model}), one finds $\chi^2_{min,VSL}=955.465$ for $\gamma=0.513$, which within the $68 \%$ confidence level makes $64<H_0<64.2$. 

Comparing the VSL model to the concordance model, it should be noted that for this value of $\gamma$ and in the redshift interval used in union 2.1 data set $0.015<z<1.414$ \citep{union},  the contribution of the quadratic term (\ref{eq:model}), is always smaller than the cosmological constant term, i.e. $0.004<\frac{F \az^{\mu\nu}\az_{\mu\nu}}{\Lambda}<0.635$. 
 In this sense, though the $\chi^2_{min,VSL}$ is higher than $\chi^2_{CM}$, it is fair to say that this model is still in the vicinity of the concordance model.  Given that the mechanism driving the late-time expansion in this model is quite different from an effective cosmological constant, we believe that the statistical proximity between the models is remarkable and suggests that other mechanisms such as a varying speed of light could be relevant to interpret the observational data. \\
 
\begin{figure}[H]
  \centering
  \includegraphics[width=0.9\linewidth]{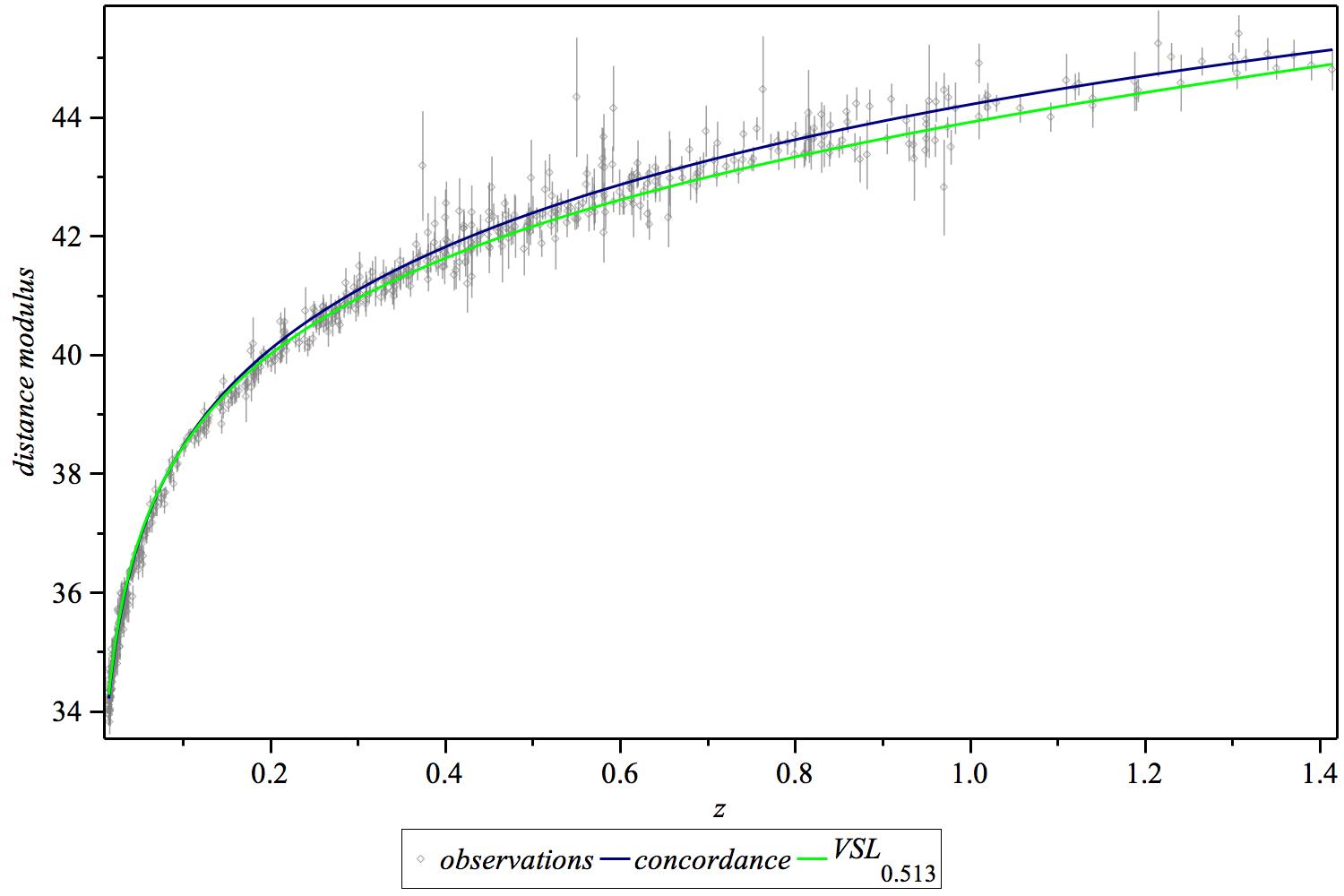}
  \caption{Distance modulus for 
  i) union 2.1 data set \citep{union}
  ii) concordance model
  iii) VSL model with $\gamma=0.513$.}
  \label{fig_dL}
\end{figure}

\begin{figure}[H]
 \centerline{\includegraphics[width=0.7\linewidth]{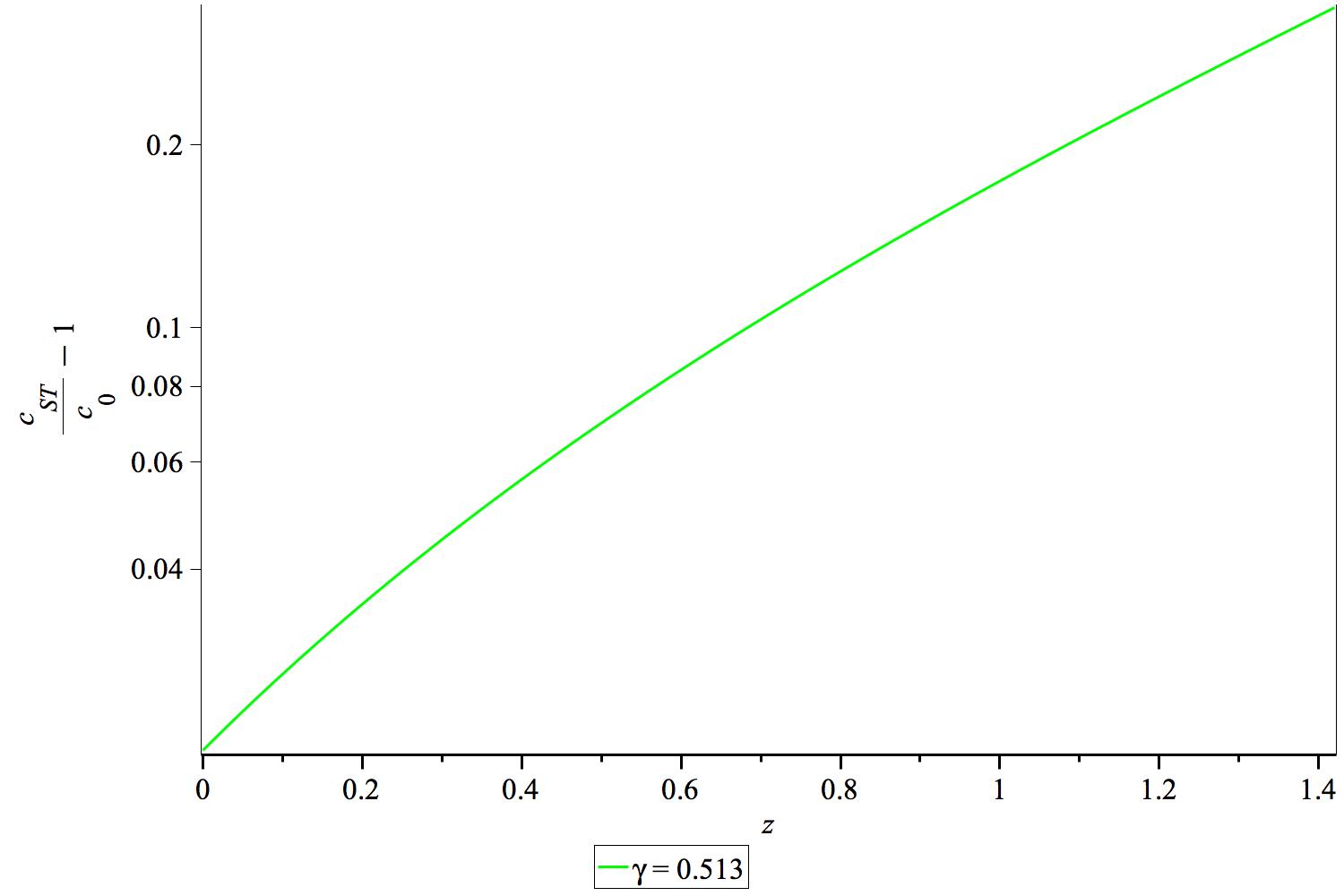}}
 \vspace*{8pt}
 \caption{\cST\ variations vs. redshift for $\gamma=0.513$.
 \protect\label{fig_cST}}
 \end{figure}

Figure \ref{fig_dL} shows the distance modulus for the VSL model with the best fit of $\gamma$, compared to the concordance model and also the union 2.1 data set.
As shown in equation \eqref{cST}, \cST\ varies with redshift for the best fit of $\gamma$.
This is shown in Figure \ref{fig_cST} .

\section{Summary and conclusions}

In this work we have retaken the debate on the different manifestations of the speed of light in gravitational scenarios. 
Considering theories formulated in both the Palatini and metric-affine approaches, where the notion of local inertial frame has some subtleties due to the independence of metric and connection, it is possible to explicitly break some of the known degeneracies. In particular, we have pointed out that for the determination of distance, which needs both measuring time and a signal going from one point to another, it is important to distinguish between \cST and $c_{EM}$. \\

Focusing on the definition of $c_{ST}$, one finds that in the case of $f(\az)$ theories $c_{ST}$ is insensitive to the form of the function $f_\az(T)$ because the two metrics are conformally related. 
In other Palatini theories, such as in the case of quadratic gravity,  $\az+\alpha \az^2+\beta \az_{\mu\nu} \az^{\mu\nu}$, or in Born-Infeld inspired gravity models, $c_{ST}$ may vary in the local frame 
due to effects of the local stress-energy density which manifest in a non-conformal way. For  inverse curvature models, this local density dependence is known to have a nontrivial impact in microscopic 
systems due to violations of the equivalence principle \cite{Olmo:2008ye,Olmo:2006zu}. \\

Due to the fact that measured distance is not a gauge covariant quantity, the definition of the distance modulus in cosmology is sensitive to the variation of \cST. We have discussed how this quantity 
should be defined in the varying speed of light case [see Eq.\eqref{newdl}], and have used that definition and the usual one to confront a certain quadratic gravity toy model with supernovae data. 
The results appear in Sec. \ref{sec:confrontation}. \\

The numerical results indicate that  the obtained $\chi^2_{CM}$ (for $\Lambda$CDM) is comparable with $\chi^2_{min,VSL}$ and $\chi^2_{min,MEdS}$. However, the MEdS model is inconsistent with Planck data, 
\citep{Planck_VFC}, within the $68\%$ confidence level.
So due to the statistical results on the MEdS model, we conclude that the local frame in which the affine connection vanishes is observationally preferred. 
Although according to the data, $\Lambda$CDM is statistically preferred over the VSL model, it should be noted that  the $\Lambda$CDM model involves the {\it ad hoc} introduction of a cosmological 
constant term (of unknown origin) which dominates the energy density of the universe, 
whereas the VSL model simply assumes a quadratic curvature correction, which could be accommodated within an effective field theory approach. 
Extension of the analysis presented here to more general gravity Lagrangians will be the subject of future work.\\

A comment regarding the effective nature of the model (\ref{eq:model}) and the unusual magnitude of the coupling constant $\Gamma$ considered here is in order. 
The relation between Palatini geometry and condensed matter physics presented in \cite{Lobo:2014nwa} indicates that in a gravitational context the matter fields can be seen as
the analogous of structural defects in crystals. As a result, the effective description
of matter at different scales necessarily leads to different types of
structural defects, which could imply a dependence of the resulting
effective dynamics on the scale.  Since different defects in different
crystals lead to different properties (such as elasticity, plasticity,
conductivity, ...) it is legitimate to admit the possibility that the
gravitational dynamics governing microscopic scales could be very different
from that governing larger scales simply because the structural defects
(or, equivalently, the effective matter fields) proper of a scale might be completely different
from those present at other scales. The kind of effective geometry (or
crystal) corresponding to a certain scale (such as an atom or the solar
system, where empty space dominates the total volume) could thus be very
different from that corresponding to cosmological models, where a
continuous fluid distribution fills all the space.  Note, in this sense,
that the precise averaging procedure to go from local scales to cosmology
is not well understood in GR, let alone in the class of theories considered
here, where the connection induces additional nonlinearities in the matter
sector. Therefore, the fact that the quadratic model considered in this
paper requires a coupling constant $F$ of an unusual magnitude does not
necessarily imply that the model could be in conflict with local gravity
experiments, where a much smaller coupling constant would be expected. In our view, at those scales the corresponding effective theory of gravity could be very different from that applicable to cosmic scales. For that reason, the viability of this model should be assessed only through its implications at the cosmological level.\\ 

Summarizing, though the statistical analysis somehow favors the  $\Lambda$CDM model, from a theoretical perspective the quadratic gravity approach is more appealing.  Due to the small difference between the $\chi^2$ of the concordance and the VSL model, it is not justified to observationally favor one model over the other. Since VSL models can affect the scenario of generating primordial perturbations and structure growth, it is important to study their compatibility with CMB and LSS observations (for more details see \cite{ruthprl,ruthprd,Moffat_VSL, Weikang}). Further research aimed at testing the VSL model with those and other observations is currently underway.

%This model is also worth exploring on the primordial perturbations, since a causal mechanism of generating primordial perturbations can be achieved by varying $c_{ST}$ cosmology in a primordial epoch. 
%This provides a very good alternative to the inflation model of gravity \cite{Moffat_VSL}.
%Also the structure growth can be used to study the compatibility of these models with CMB and LSS \cite{Weikang}.

\section*{Acknowledgments}
The authors are grateful to Stefan Czesla for his helpful discussions.
A. Izadi is supported by K. N. Toosi University of Technology and is also thankful for the hospitality of people at the observatory of Hamburg. G. J. Olmo is supported by a Ramon y Cajal contract and the Spanish grant FIS2014-57387-C3-1-P (MINECO/FEDER, EU). Support from  the Consolider Program CPANPHY-1205388, the Severo Ochoa Grant SEV-2014-0398 (Spain), the CNPq project No. 301137/2014-5 (Brazilian agency), and Red Tem\'{a}tica de Relatividad y Gravitaci\'{o}n FIS2016-81770-REDT (MINECO/FEDER, EU) is also acknowledged.  
This paper is based upon work from COST Action CA15117, supported by COST (European Cooperation in Science and Technology).

\bibliographystyle{spphys}
\bibliography{bibliography}
\end{document}